\documentclass{an}
\usepackage{graphicx}
\usepackage{times}
\Volume{00}              % issue in current year, format {NN}
\Year{0000}              % format: {YYYY}
\Month{00}               % format: {M} or {MM} (number), month of print version
\Pagespan{000}{000}      % format: {start page}{last page}

\begin{document}
\lhead[\thepage]{A.N. Author: Title}
\rhead[Astron. Nachr./AN~{\bf XXX} (200X) X]{\thepage}
\headnote{Astron. Nachr./AN {\bf 32X} (200X) X, XXX--XXX}

\title{The CYDER Survey: First Results}

\author{Francisco J. Castander\inst{1,2,3}, Ezequiel
Treister\inst{2,3}, Jos\'e Maza\inst{2}, Paolo S. Coppi\inst{3},
Thomas J. Maccarone\inst{4}, Stephen E. Zepf\inst{5}, Rafael
Guzm\'an\inst{6}, Mar\'{\i}a Teresa Ruiz\inst{2}}
\institute{Institut d'Estudis Espacials de Catalunya/CSIC, Gran
Capit\`a 2-4, 0834 Barcelona, Spain
\and
Departamento de Astronom\'{\i}a, Universidad de Chile, Casilla 36-D, Santiago,
Chile
\and
Astronomy Department, Yale University, P.O. Box 208101, New
Haven, CT06520, USA
\and
SISSA, via Beirut 2-4, 34014 Trieste,
Italy
\and
Department of Physics and Astronomy, Michigan State
University, East Lansing, MI 48824, USA
\and
Department of Astronomy,
University of Florida, P.O. Box 112055, Gainesville, FL 32611, USA}

\date{Received {date will be inserted by the editor}; accepted {date
will be inserted by the editor}}

\abstract{We present the Cal\'an-Yale Deep Extragalactic Research
(CYDER) Survey. The broad goals of the survey are the study of stellar
populations, the star formation history of the universe and the
formation and evolution of galaxies. The fields studied include
Chandra deep pointings in order to characterize the X-ray faint
populations. Here we present the results on the first fields
studied. We find that the redshift distribution is consistent with
that found in the Chandra Deep Field North. The distribution of
hardness ratios is, however, softer in our sample. We find a high
redshift quasar, CXOCY J125304.0-090737 at $z=4.179$, which suggests that
the abundance of low luminosity high redshift quasars may be larger
than what would be expected from reasonable extrapolations from the
quasar optical luminosity function.  \keywords{surveys  ---  quasars: general --- galaxies: active --- X-rays  --- galaxies: evolution} } 

\correspondence{fjc@ieec.fcr.es}

\maketitle

\section{Introduction: the CYDER Survey}
The Cal\'an-Yale Deep Extragalactic Research (CYDER) Survey is a
collaborative effort between the Universidad de Chile and Yale
University to study in detail faint stellar and extragalactic
populations in survey mode. The broad scientific goals are directed
towards the core observing goals of the new generation of large
optical and millimeter facilities. The program takes full advantage of
these facilities by combining deep optical and near-IR photometric and
spectroscopic observations on wide field cameras and spectrographs
using a wide variety of 4-m and 8-m class telescopes.

\subsection{Strategy}
Fields were selected at high galactic latitude to minimize the effects
of extinction. They were also spread out in right ascension to allow
flexibility in the allocation of telescope time. Amongst our fields,
we selected fields observed by the Chandra X-ray Observatory satellite
with exposure times longer than 50 ks.

The CYDER survey original design goal was to cover 1 square degree
down to limiting magnitudes $U\sim26$, $B\sim26.5$, $V\sim26$,
$R\sim25.5$, $I\sim25$, $z\sim24$, $J\sim22$ and $K\sim20$ at ${\rm
S/N} \sim 10$.  So far, the optical coverage is larger than 1 square
degree in some filters while there is no area coverage in others. In
the near infrared only 1/2 of a square degree has been imaged.
Optical spectroscopy is underway in a few selected fields, while near
infrared spectroscopy has not started yet.

\section{Observations}
The first fields studied were three of the earliest deep Chandra
pointings to become publicly available. One of these field is in the
Northern hemisphere. It is the Chandra pointing towards the blazar
1156+295. The X-ray exposure time was 75ks.  The other two fields are
in the South. They were pointing to the Hickson compact group HCG62
(exposure time 50ks) and the blazar Q1127-145 (30ks). These two
Southern fields will be the ones discussed in this paper.

\subsection{X-ray data}
Both the HCG62 and Q1127-145 fields were observed with the Chandra
ACIS-S instrument. We retrieved these images from the archive and
analyzed them using standard techniques with the CIAO package. In the
HCG62 field we detect 34 X-ray sources in the s3 ACIS CCD and 16
sources in the s4. In Chandra pointing towards Q1127-145, the s3 CCD was
the only CCD read. It was read in subraster mode and therefore only 5
X-ray sources are detected in that field.

\subsection{Optical and Infrared Imaging}
We have observed both X-ray fields with the CTIO 4m MOSAIC-II
camera. The total integration time for the HCG62 field is 200 minutes
in U, 36min in B, 80min in V and 25 min in I under 1.0-1.5''
conditions. In the Q1127-145 field the integration times are 80, 75 and
25 minutes in V, R and I respectively. Images were reduced using
standard techniques with the IRAF/MSCRED package.

In the near infrared we have observed both fields at Las Campanas
Observatory with the DuPont 2.5m telescope using the Wide Field
InfraRed Camera during 60 minutes in J and 120 minutes in K$_s$ under
typical 0.6-0.7'' seeing conditions. We have reduced the data with the
IRAF DIMSUM package following standard procedures.

\subsection{Optical Spectroscopy}
Follow up spectroscopy of these fields was obtained at the ESO Cerro
Paranal Observatory with the UT4/Yepun telescope using the FORS2/MXU
instrument and at the Las Campanas Observatory with the LDSS-2
instrument at the Magellan Baade telescope. Several masks were
designed which included slits for most, but not all, of the X-ray
sources in these fields. Masks were observed for approximately two
hours. The instrument configuration used resulted in a spectral
resolution of $R\sim520$ at VLT and $R\sim350$ at Magellan. The
spectra were reduced using standard techniques with IRAF and
calibrated in wavelength using He-Ar comparison lamps exposures and
the night sky lines.

\section{First Results}
In our first two Southern fields we have spectroscopically identified
25 X-ray sources which corresponds to approximately half of the total
X-ray sources detected. Table~\ref{tab1} summarizes the percentages of
the different type of objects in our sample. For comparison we also
present the results for the Chandra Deep Field North (CDF-N; Brandt et
al 2001, Barger et al 2002) and the Chandra Multiwavelength Project
(ChaMP; Green et al 2003). We have grouped the different object types
in broad classes and have translated the source types of Barger et al
(2002) into these types.  The CDF-N, CYDER and ChaMP surveys reach
different X-ray flux limits; the CDF-N survey reaching the faintest,
the ChaMP survey, the brightest. Table~\ref{tab1} demonstrates how the
source composition changes with flux limit in an X-ray survey,
although the identification incompleteness may hide or enhance certain
trends. At bright flux limits, the broad line active galactic
nuclei/quasars dominate the extragalactic sources. Their percentage
contribution diminishes as the flux limit lowers due to the appearence
of a new population of X-ray fainter narrow emission lines AGN/QSOs
and normal galaxies.

\begin{table}[h]
\caption{Approximate percentages of different type of sources in the
CDF-N (Barger et al 2002), the CYDER and ChaMP (Green et al 2003) surveys}
\label{tab1}
\begin{tabular}{lccc}\hline
& CDFN & CYDER & ChaMP\\ 
\hline
Stars               &  5\%  &  5\% & 10\%  \\
Broad line AGN/QSO  & 25\%  & 40\% & 65\%  \\
Narrow line AGN/QSO & 50\%  & 40\% & 15\%  \\
Galaxies            & 20\%  & 15\% & 10\%  \\
\hline
Total Number of Identified Sources  & 170   & 25   & $\sim$200 \\
Total Number of Detected Sources    & 370   & 55   & - \\
\end{tabular}
\end{table}

\begin{figure}
\resizebox{\hsize}{!}  {\includegraphics[]{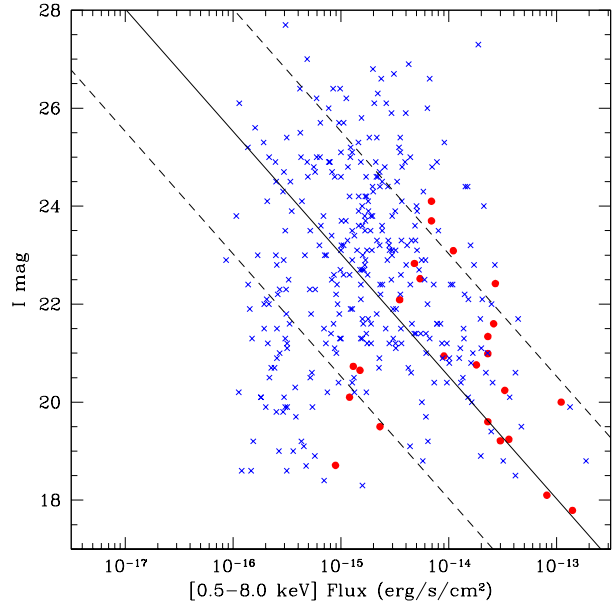}}
\caption{$I$ magnitude plotted against full band (0.5-8.0 keV) X-ray
flux. Blue crosses represent the CDF-N sources and red circles, the
CYDER sources. The black solid line represents the line of constant
X-ray-to-optical flux ratio, that is, objects with $f_I/f_x=1$, where
$f_I$ is the flux in the I optical band and $f_x$ the X-ray flux. The
two dashed lines correspond to objects with ${\rm log}
(f_I/f_x)=\pm1$, enclosing the region populated by {\it normal}
AGN/QSOs}
\label{fig1}
\end{figure}

Figure~\ref{fig1} shows the I band and total X-ray flux of our Chandra
sources. For comparison we also plot the CDF-N sources. In our sample
we find that approximately 30\% of our sources do not show an optical
counterpart down to $I\sim24$.  In the case of the CDF-N 16\% are
undetected down to $I\sim26$ and 35\% down to $I\sim24$. Overall, the
distribution of our sources in the I .vs. X-ray flux plane occupy the
same parameter space as the CDF-N sources cut at our same X-ray flux
limit.

\begin{figure}
\resizebox{\hsize}{!}  {\includegraphics[]{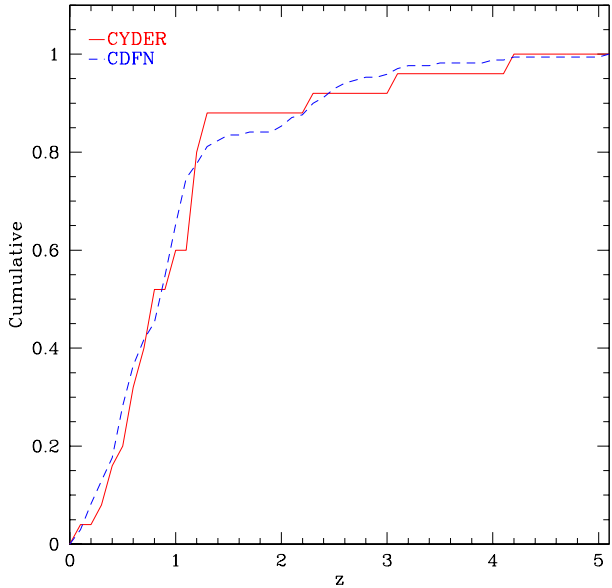}}
\caption{Cumulative redshift distribution of CYDER (solid red) and CDF-N
(dashed blue) sources.}
\label{fig2}
\end{figure}

We compare the CYDER and CDF-N source redshift distributions in
Figure~\ref{fig2}. A Kolmogorov-Smirnov (KS) test shows that both
distribution are compatible with having been drawn from the same
parent population and are therefore statistically indistinguishable.
Given that the CDF-N sources are typically fainter, we have cut their
sample at the effective X-ray flux limit of our sample. If we compare
the redshift distribution in this case, they remain to be
statistically compatible. It is worth noting that in our sample there
are 5 QSOs at a redshift $z\sim1.2$ in the same field. This large
scale structure feature is noticeable in our redshift distribution and
stresses the need to study sufficient sources and fields as to not be
biased by such structures.

We have also compared the X-ray properties of our sample to the CDF-N
sources. We find that our sources are in general softer than those in
the CDF-N (see Treister \& Castander 2003). A KS test indicates that
the hardness ratio distributions are incompatible with being drawn
from the same parent population. This results may be expected as the
CDF-N sources are typically fainter and fainter sources are in general
harder (e.g., Giacconi et al 2001; Tozzi et al 2001; Brandt et al
2001; Stern 2002). However, if we impose our effective flux limit to
the CDF-N sample we still find that our sources are significantly
softer than the reduced brighter CDF-N subsample.

\begin{figure}
\resizebox{\hsize}{!}  {\includegraphics[angle=270]{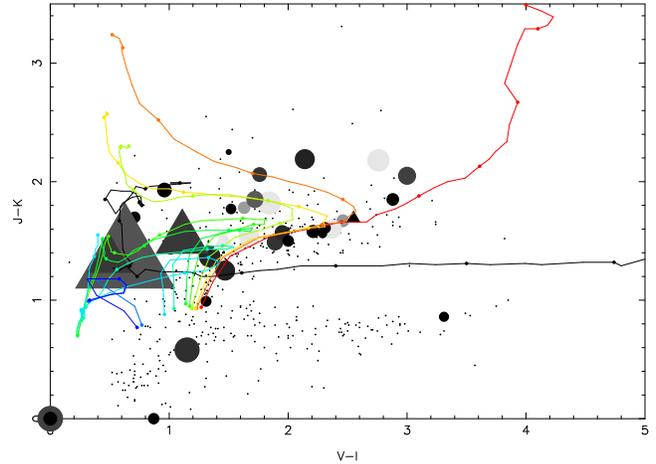}}
%{\includegraphics[angle=270]{ccJKvsVI.ps}}
\caption{$J-K$ .vs. $V-I$ colour-colour plot including the CYDER X-ray
sources. The small dots represent the colours of the objects in a
reduced area of our optical and near infrared coverage of our Chandra
fields. The main sequence stars (band of objects from $J-K\sim0.3$ and
$V-I\sim0.6$ to $J-K\sim0.8$ and $V-I\sim3.5$) nicely separate from
the rest of the extragalactic sources. The grey symbols indicate the
X-ray sources (circles: sources in the HCG62 field; triangles:
Q1127-145 field). The size of the circles and triangles is
proportional to the X-ray flux. The grey intensity of the symbols
shows the hardness ratio. The softest sources (HR=-1) are represented
in dark grey and the hardest sources (HR=1) in light grey. The black
line indicates the expected location of a QSO at different
redshifts. The small dots in the line indicate separations of 0.5
units in redshift. The track starts at $z=0$ at $J-K\sim1.9$ and
$V-I\sim1.1$, then moves blue-wards in $V-I$ at approximately constant
$J-K$ up to $z\sim2$, then moves blue-wards in $J-K$ at approximately
constant $V-I$ up to $z\sim3-3.5$ and finally quickly moves red-wards
in $V-I$ at approximately constant $J-K$. The grey (coloured in
electronic version) tracks represent the expected evolutionary path of
different galaxy types. The dark grey curve most to the right (red) is
a 1 Gyr burst of star formation occurring 15 Gyr ago that evolves
passively since. The other evolutionary tracks are computed with star
formation prescriptions that broadly reproduce the observed
photometric properties of E, S0, Sa, Sb, Sbc, Sc, Sd and irregular
galaxies (from right to left run from earlier types to later
types). The points at the bluest $J-K$ colour in each track represent
those galaxies at $z=0$. Small dots on the tracks are then spaced in
0.5 units in redshift.}
\label{fig3}
\end{figure}

We also investigate what the optical properties of our sample
are. Figure~\ref{fig3} is a colour-colour plot including our X-ray
sources counterparts (see the figure caption for an explanation of the
symbols). Two sources lie on the stellar locus. One is
spectroscopically identified as a star, while the other has not been
observed spectroscopically. The rest of the sources populate the
region of colour-colour space of extragalactic sources. Some of our
broad emission line AGN/QSOs are close to the expected location of
typical quasars. Others, on the other hand, deviate from their
expected position indicating that they are probably reddened. The
majority (although not all) of our soft sources lie close to the
expected locus of early-type galaxies. The hardest sources populate
the regions of mildly active galaxies at redshifts
$z\sim0.5-1.0$. They may be at such locations because they are indeed
this type of galaxies or because they have been reddened to that part
of colour-colour space.

We compare the photometric and X-ray properties of our sources. We
find that the hardest sources are preferentially redder than the rest
of the objects. However, we also find soft sources that are red,
implying that while blue sources are preferentially soft, red
sources can be either X-ray hard or soft. Given the reduced number of
sources in our sample, this effect could simply be a statistical
fluctuation.

The spectra of our sources reveal a diverse variety of type of
objects.  We find typical examples of broad and narrow emission line
AGN/QSOs. We find typical old stellar population spectral energy
distributions. There are also examples of poststarburst galaxies and
objects whose spectral classification is difficult as they have broad
emission lines typical of quasars and spectral characteristics of old
stellar population with some moderate on-going star formation. Some of
our objects show obvious signs of strong absorption.

\begin{figure}
\resizebox{\hsize}{!}  {\includegraphics[]{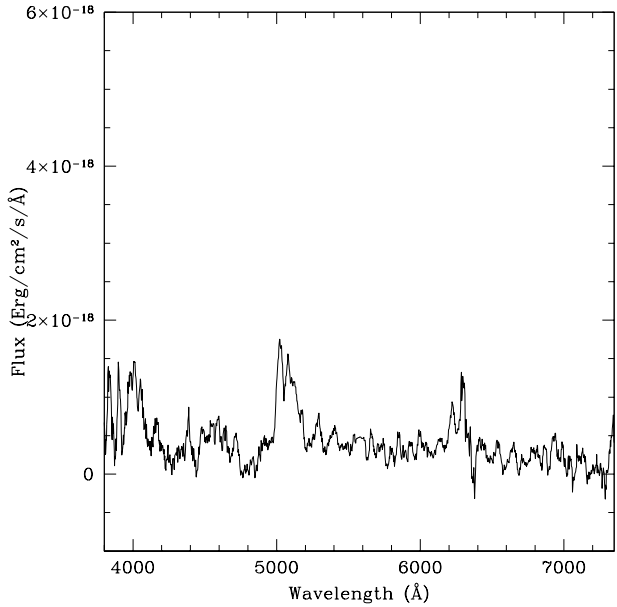}}
\caption{Optical spectrum of CXOCY J125241.0-091622 obtained at
UT4/Yepun VLT.}
\label{fig4}
\end{figure}

Here we comment on two of our X-ray sources.
CXOCY J125241.0-091622 is a quasar at redshift $z=2.282$
(Figure~\ref{fig4}) with a harder than usual X-ray spectrum
$\Gamma\sim1.3$. In the optical its $V-K$ color is 4.0 which is one
magnitude redder than the expected colour of a prototype quasar at
this redshift (Figure~\ref{fig3}). Both X-ray and optical data thus
indicate that this is an obscured object. Such objects are predicted
in models as contributors to the X-ray Background at faint fluxes.

\begin{figure}
\resizebox{\hsize}{!}  {\includegraphics[angle=270]{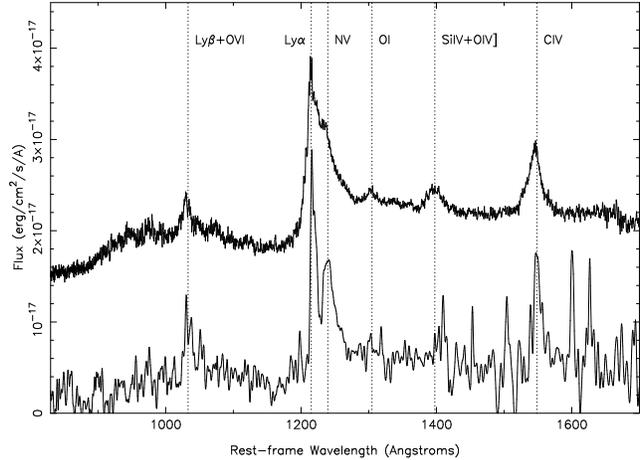}}
\caption{Optical spectrum of CXOCY J125304.0-090737 obtained at
UT4/Yepun VLT. For comparison, we also show the error weighted average
of the SDSS Early Data Release QSO spectra at $z>4$. This composite
spectrum have arbitrary zero point and scaling offsets for display
purposes and it is shown above the spectrum of CXOCY
J125304.0-090737. The most common QSO emission lines are indicated as
dotted lines. The emission lines in CXOCY J125304.0-090737 are
narrower than the typical SDSS spectrum.}
\label{fig5}
\end{figure}

CXOCY J125304.0-090737 is an optically faint quasar
($M_B=-23.69+5\times log (h_{65})$) with a typical QSO spectrum
(figure~\ref{fig5}). In X-rays, CXOCY J125304.0-090737 is also X-ray
faint ($f_X = 1.7\pm 0.4 \times 10^{-15}$ ergs s$^{-1}$ cm$^{-2}$ in
the [0.5-2.0] keV band) with a somewhat harder spectrum
($\Gamma\sim1.7$ or $HR\sim-0.35$) than typical low redshift or high
redshift optically selected quasars. We speculate that a reflection
component can slightly harden the spectrum but by no means is this the
only mechanism. CXOCY J125304.0-090737 X-ray-to-optical emission is
X-ray strong ($\alpha_{ox}\sim-1.35$) compared to high
redshift optically selected quasars (see Castander et al 2002 for
further details).

CXOCY J125304.0-090737 is the only quasar above $z=4$ found so far in
the CYDER survey. However, the space density implied by its discovery
is higher than reasonable extrapolations of the quasar optical
luminosity function to fainter luminosities at high redshifts (Fan et
al 2001) . Although this object by itself does not constrain the
faint end of the luminosity function at high redshift, it demonstrates
the possibilities that X-ray surveys have to achieve this goal.

\section{Summary}

We have briefly presented the CYDER survey, which is currently
underway. We have focused in the first fields studied that were
chosen to coincide with moderately deep Chandra X-ray pointings. We
have investigated the nature of the X-ray sources in these fields. We
find that the broad X-ray and optical properties of our sources are
similar to the ones studied in the CDF-N. The only difference is that
our sources appear to be softer. We also stress the fact that our
X-ray sources are of diverse optical spectral types and have
highlighted the discovery of a high redshift X-ray selected quasar.

The current X-ray surveys (most of which are presented in this volume)
that have flourished with the launch of the new X-ray observatories
promise to be key to our understanding of the faint X-ray
populations. The CYDER survey will be one of the surveys contributing
to pin down the evolution of accretion material on to black holes
which seems to be closely related to the star formation history of
the universe.

\acknowledgements

We acknowledge the financial support received from the Fundaci\'on
Andes that has enabled the collaboration between the Universidad de
Chile and Yale University and therefore has made the CYDER survey possible.

%\begin{appendix}
%\section{}
%\end{appendix}


\begin{thebibliography}{}
\bibitem{} Barger, A. J. et al.: 2002, AJ, 124, 1839
\bibitem{} Brandt, W. N. et al.: 2001, AJ, 122, 2810
\bibitem{} Castander, F. J. et al.: 2002, AJ, submitted
\bibitem{} Giacconi, R. et al.: 2001, ApJ, 551, 624
\bibitem{} Green, P.J. et al: 2003: AN, this volume
\bibitem{} Fan, X. et al: 2001, AJ, 121, 31
\bibitem{} Stern, D. et al.: 2002, AJ, 123, 2223 
\bibitem{} Tozzi, P. et al.: 2001, ApJ, 562, 42
\bibitem{} Treister, E. \& Castander, F. J.: 2003, AN, this volume
\end{thebibliography}
\end{document}